\documentclass[aps,prl,twocolumn,preprintnumbers,superscriptaddress,nobibnotes,floatfix,longbibliography,nofootinbib]{revtex4-1}

\pdfoutput=1
\usepackage{amsmath,amsfonts,amssymb,mathrsfs,graphicx,color,longtable}
\usepackage{hyperref}
\usepackage{graphicx}
\usepackage{amsfonts,amsmath,amssymb,bm}
\usepackage{array}
\usepackage{color}
\usepackage{enumitem}
\usepackage[explicit]{titlesec}
\usepackage[utf8]{inputenc}
\usepackage[normalem]{ulem}

\usepackage{feynmf}

\hypersetup{
     colorlinks   = true,
     citecolor    = red,
     urlcolor     = blue,
     linkcolor    = blue
}

\usepackage{xcolor}

\newcommand{\be}{\begin{equation}}
\newcommand{\ee}{\end{equation}}

\newcommand{\PRL}{Phys. Rev. Lett.}

\graphicspath{{figs/}}

\newcommand\Sfactor{{\mathcal{S}}}

\makeatletter

\def\hhref#1{\href{http://arxiv.org/abs/#1}{arXiv:#1}}
\usepackage{xstring} 
\newcommand{\hhrefq}[1]{\IfSubStr{#1}{:}{\href{http://inspirehep.net/search?ln=en&ln=en&p=#1&of=hb&action_search=Search&sf=&so=d&rm=&rg=25&sc=0}{InSpires:#1}}{\hhref{#1}}}

\def\art{\@ifnextchar[{\eart}{\oart}}
\def\eart[#1]#2#3#4#5#6{{\rm #2}, {\em #3 \bf #4} {\rm (#6) #5} ({\em #1})}
\def\article{\@ifnextchar[{\earticle}{\oarticle}}
\def\oarticle#1#2#3#4#5#6{{\rm #1}, {``#6''}, {\rm #2 #3 (#5) #4}}
\def\earticle[#1]#2#3#4#5#6#7{{\rm #2}, {``#7''}, {\rm #3 #4 (#6) #5}  [\hhrefq{#1}]}
\def\hepart[#1]#2{{\rm #2, \sl#1}}
\def\heparticle[#1]#2#3{#2, { ``#3''} [\hhrefq{#1}]}

\begin{document}

\hspace{2.1in}
MIT-CTP/5284

\vspace{0.5cm}

\title{Accidentally Asymmetric Dark Matter}

\author{Pouya Asadi}
\thanks{{\scriptsize Email}: \href{mailto:pasadi@mit.edu}{pasadi@mit.edu}; 
}
\affiliation{Center for Theoretical Physics, Massachusetts Institute of Technology, \\ Cambridge, MA 02139, USA.}
\author{Eric David Kramer}
\thanks{{\scriptsize Email}: \href{mailto:david.kramer@mail.huji.ac.il}{david.kramer@mail.huji.ac.il}; 
}
\affiliation{Racah Institute of Physics, Hebrew University of Jerusalem, Jerusalem 91904, Israel.}
\author{Eric Kuflik}
\thanks{{\scriptsize Email}: \href{mailto:eric.kuflik@mail.huji.ac.il}{eric.kuflik@mail.huji.ac.il}; 
}
\affiliation{Racah Institute of Physics, Hebrew University of Jerusalem, Jerusalem 91904, Israel.}
\author{Gregory~W.~Ridgway}
\thanks{{\scriptsize Email}: \href{mailto:gridgway@mit.edu}{gridgway@mit.edu}; 
}
\affiliation{Center for Theoretical Physics, Massachusetts Institute of Technology, \\ Cambridge, MA 02139, USA.}
\author{Tracy R. Slatyer}
\thanks{{\scriptsize Email}: \href{mailto:tslatyer@mit.edu}{tslatyer@mit.edu}; 
}
\affiliation{Center for Theoretical Physics, Massachusetts Institute of Technology, \\ Cambridge, MA 02139, USA.}
\author{Juri Smirnov}
\thanks{{\scriptsize Email}: \href{mailto:smirnov.9@osu.edu}{smirnov.9@osu.edu}; 
}
\affiliation{Center for Cosmology and AstroParticle Physics (CCAPP), The Ohio State University, Columbus, OH 43210, USA}
\affiliation{Department of Physics, The Ohio State University, Columbus, OH 43210, USA}

\date{\today}

\begin{abstract}
We study the effect of a first-order phase transition in a confining $SU(N)$ dark sector with heavy dark quarks. The baryons of this sector are the dark matter candidate. 
During the confinement phase transition the heavy quarks are trapped inside isolated, contracting pockets of the deconfined phase, giving rise to a second stage of annihilation that dramatically suppresses the dark quark abundance. 
The surviving abundance is determined by the local accidental asymmetry in each pocket.
The correct dark matter abundance is obtained for $\mathcal{O}(1-100)$ PeV dark quarks, above the usual unitarity bound. 
\end{abstract}

\maketitle

{\bf Introduction.---}
\label{sec:intro}
Despite making up $85\%$ of the cosmic matter abundance, the particle nature of dark matter (DM) is still unknown. One fundamental question is whether dark matter's constituents are elementary particles or composite objects. Many studies in the literature have considered the possibility of the DM being a composite state of a confining dark gauge group \cite{hep-ph/0608055,0903.3945,0906.0577,0907.1007,0909.2034,1003.4729,1005.0008,1102.0282,1106.3101,1109.3513,1204.2808,1209.6054,1301.1693,1307.2647,1312.3325,1402.3629,1408.6532,1411.3727,1503.04203,1503.08749,1602.00714,1604.04627,1606.00159,1707.05380,1801.01135,1802.07720,1811.06975,1811.08418,1904.12013,1905.08810,1911.04502,2004.03299,2006.16429,2008.10607,2010.13678}.

In such scenarios, two events in cosmic history can influence the relic abundance: the freezeout of the interactions that set the constituent quark abundance and the phase transition that converts elementary constituents into composite states. In the case that the confining phase transition happens prior to the freezeout, the details of the phase transition process are irrelevant for the final relic abundance \cite{1503.08749}. 
The DM mass in this case is expected to approach the maximum value allowed by unitarity, i.e. $m_{\rm DM} \sim 100 \, \rm TeV$ \cite{GriestKamionkowski,1407.7874,1904.11503}. 

In this letter we focus on the opposite regime, where the phase transition happens at much lower temperatures than the freezeout of constituent quarks. This regime has been considered in Refs.~\cite{1707.05380,1801.01135,1802.07720,1811.08418,1905.08810}; however, the detailed dynamics of the phase transition have never been taken into account. We show in this work that these details can significantly affect the DM abundance calculation. In this regime the phase transition is strongly first-order and features complex bubble dynamics. We study the effect of these bubbles on the evolution of quarks and bound states. We find that in a large fraction of the parameter space, the relic abundance is strongly affected by the bubble dynamics; the dark quarks are compressed within contracting pockets of the deconfined phase, leading to a second stage of efficient annihilation. 

In this paper we summarize our main findings about the effect of this first-order phase transition on the DM relic abundance, while more details are provided in a companion paper~\cite{bigpaper}.

{\bf Thermal History.---}
The high energy Lagrangian of our model consists of a dark non-abelian $SU(3)$\footnote{Our results can straightforwardly be extended to the case of $SU(N\geqslant 3)$ or more flavors of sufficiently-heavy quarks, as long as the theory remains asymptotically free. We also assume the CP-violating $\theta$ angle for this gauge group is zero.} gauge group, and a single flavor of vector-like, fundamental fermions with an explicit mass term:
\begin{align}
\mathcal{L} \supset -\frac{1}{4} G^{\mu \nu} G_{\mu \nu} +  \bar{q} \left( i \gamma_\mu D^\mu -  m_{q} \right)  q\,,
\end{align}
where $G^{\mu\nu}$ is the dark $SU(3)$'s field strength, $\Lambda$ is its confinement scale, and $m_q$ is the dark quark mass. We focus on the range of parameters $\Lambda \lesssim 0.01 m_q$. Below this scale, quarks and gluons are all confined inside glueballs, mesons, or baryons. The stable baryons are the DM candidate in this setup.

We assume this sector is connected to the SM through an unspecified portal. The portal should keep the two sectors in thermal equilibrium, enable the decay of the glueballs and the mesons into the SM, and respect the baryon number that stabilizes our DM candidate. 

Since $m_{q} \gg \Lambda$, the dark quark abundance freezes out before the phase transition takes place. Just before the onset of the phase transition, which occurs at a critical temperature of $T_c=\Lambda$, the only abundant degrees of freedom are dark gluons. 
For such a large quark mass, the confining transition is similar to that of a pure Yang-Mills theory and is of first-order~\cite{Svetitsky:1982gs,Kaczmarek:1999mm,Alexandrou:1998wv,Aoki:2006we,Saito:2011fs}. 

\begin{figure}
\resizebox{\columnwidth}{!}{
\includegraphics[scale=1]{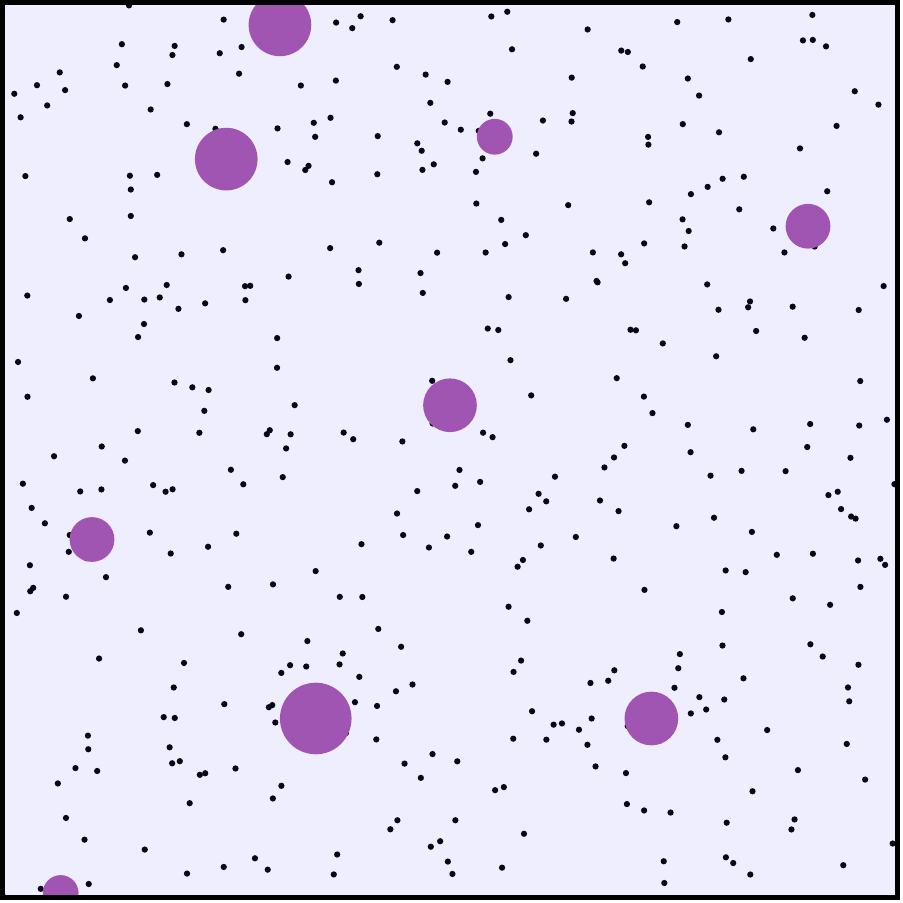}
\includegraphics[scale=1]{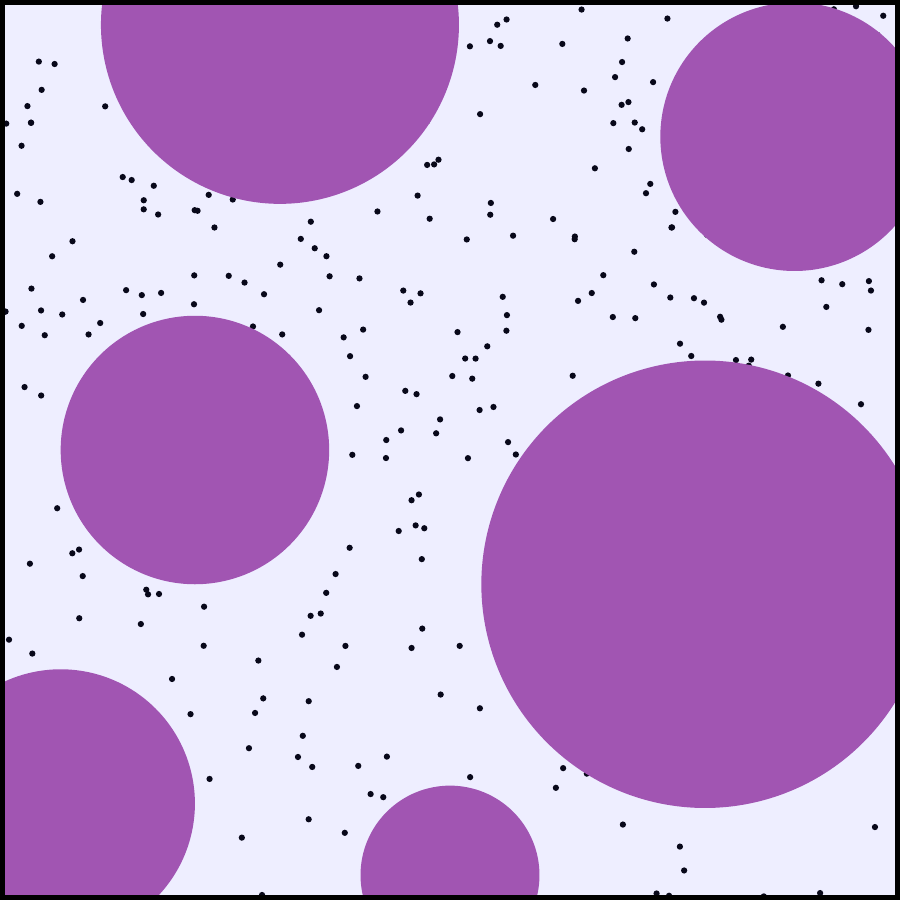}
\includegraphics[scale=1]{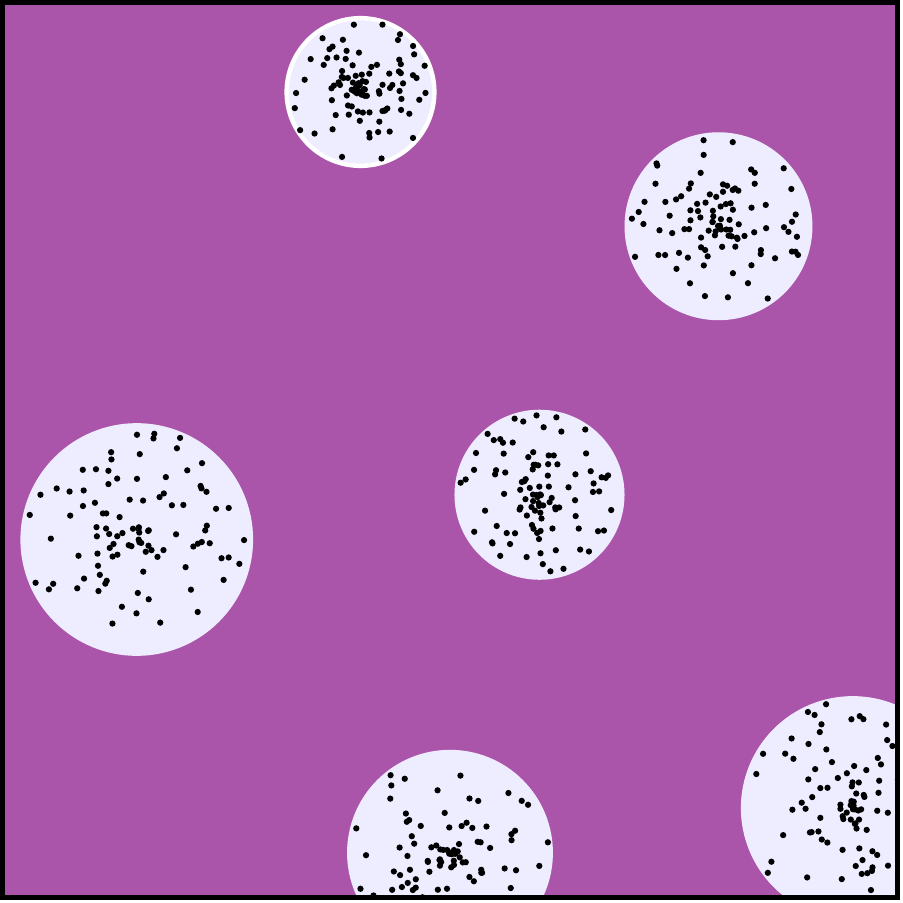}
}\\
\resizebox{\columnwidth}{!}{
\includegraphics[scale=1]{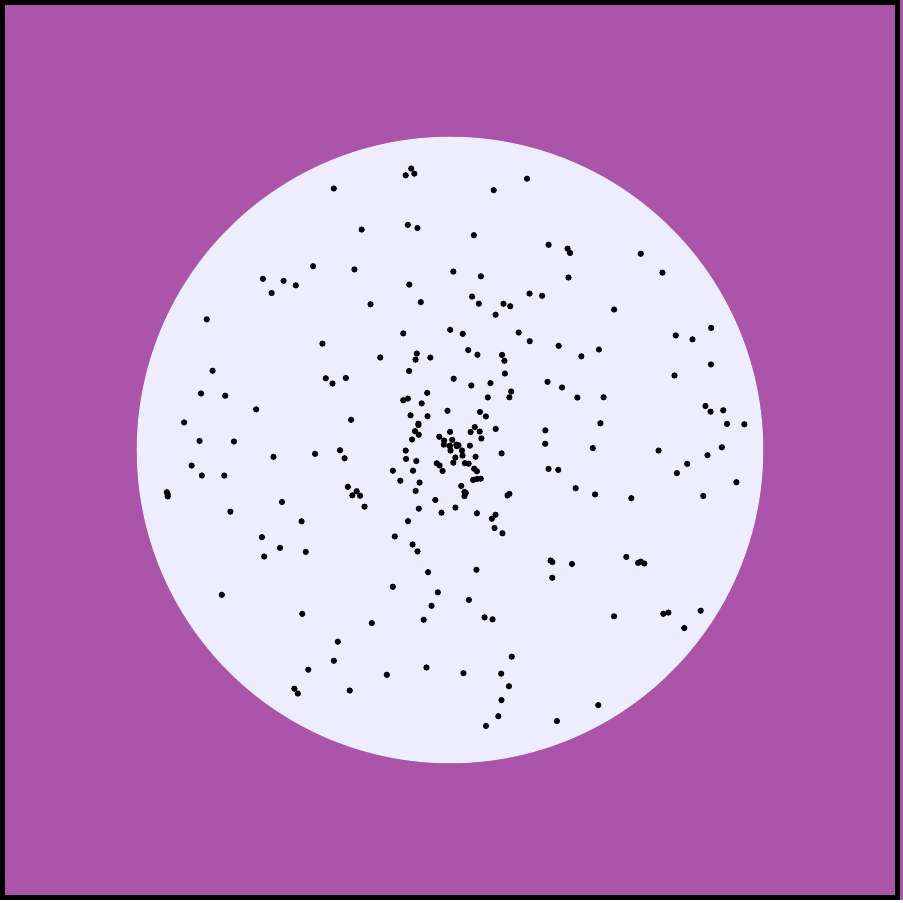}
\includegraphics[scale=1]{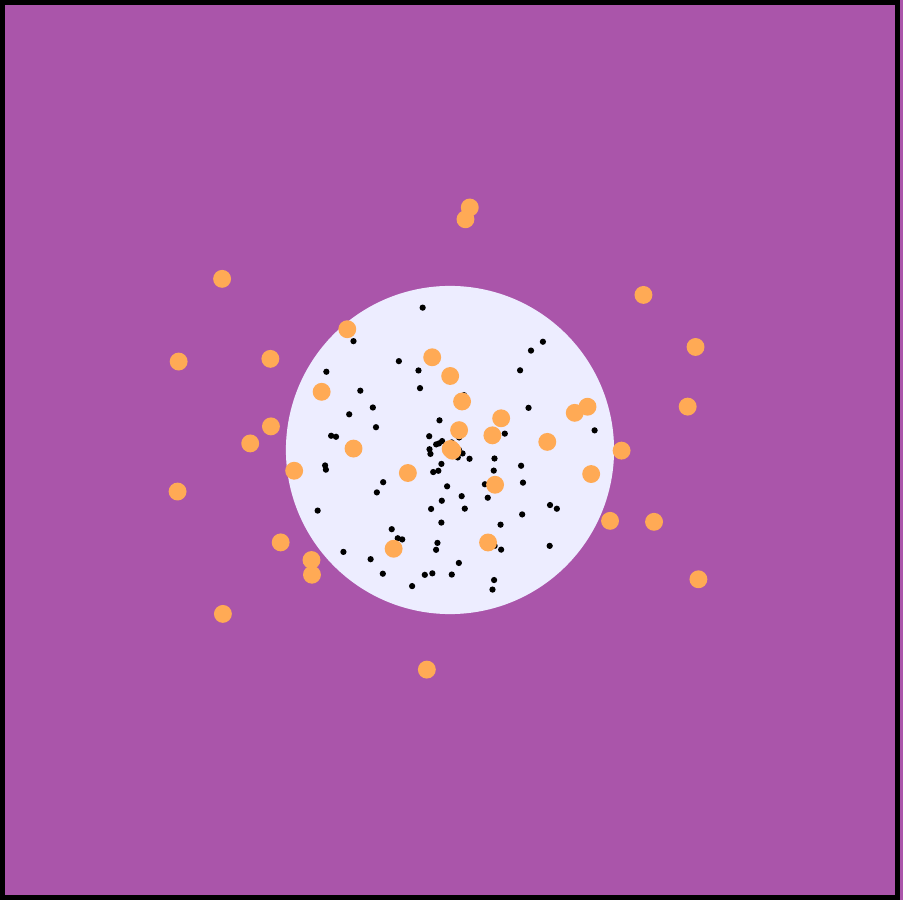}
\includegraphics[scale=1]{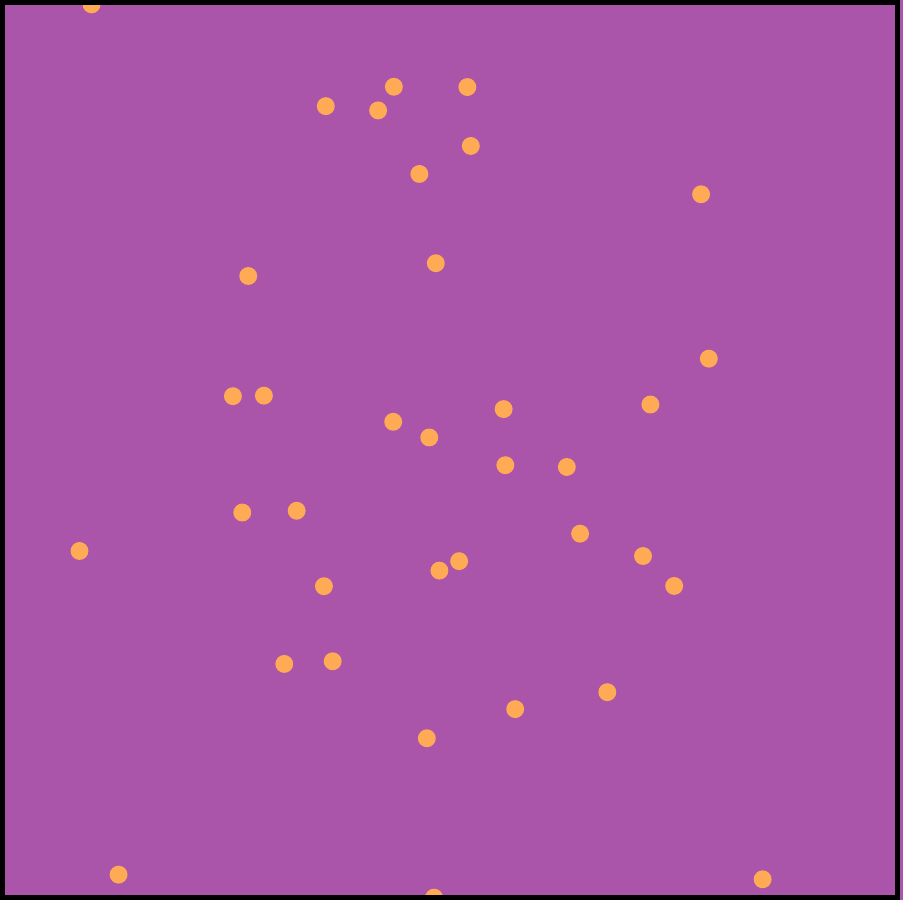}
}
\caption{Different stages of the phase transition. \textbf{Top-Left:} once the universe cools down to slightly below $T_c$, the bubbles of the confined phase (purple regions) start nucleating in a sea of the deconfined phase (light blue regions). \textbf{Top-Middle:} the nucleated bubbles keep growing until they start running into each other and percolating to make larger bubbles. \textbf{Top-Right:} eventually most of the universe converts to the confined phase, with small isolated pockets of the deconfined phase. Since isolated quarks (black dots) cannot move into the confined phase, all the quarks, which are up to this point well-separated, are gathered inside these ever-contracting pockets. \textbf{Bottom-Left:} a single contracting pocket. The quarks inside it are initially well-separated from one another. \textbf{Bottom-Middle:} as the pocket contracts, the free quarks start annihilating or forming bound states. The color-neutral bound states (orange dots) can move into the confined phase region. \textbf{Bottom-Right:} eventually the pockets disappear. A substantial fraction of quarks annihilate away during the contraction, with the residue surviving in the form of color-neutral baryons that comprise the DM today.
}
\label{fig:phases}
\end{figure}

Once the deconfined plasma supercools to slightly below $T_c$, bubbles of the confined phase start nucleating. Following Ref.~\cite{Witten:1984rs}, in Fig.~\ref{fig:phases} we schematically show how this phase transition proceeds. Initially bubbles nucleate and grow in isolation; once an $\mathcal{O}(1)$ fraction of the universe converts to the confined phase, the bubbles percolate and form a confined-phase sea surrounding isolated pockets of the deconfined phase. These pockets contract until they vanish.

Lattice studies show that the potential between two quarks in the deconfined phase flattens as they are moved away from each other \cite{Kaczmarek:1999mm}. In the confined phase, however, this potential increases indefinitely with the quark separation. As a result, the energy cost of moving a quark into a confined phase bubble is very large. 
One might suppose that this large energy could spontaneously convert to a quark -- anti-quark pair, allowing the original quark to enter the bubble within a meson. However, the large ratio $m_q/\Lambda$ severely suppresses the pair production rate~\cite{0805.4642}. Thus, the quarks are all trapped in the deconfined phase pockets.

As the pockets contract, they compress the free quarks inside them. This quark density enhancement leads to an eventual recoupling of the annihilation and bound-state-formation reactions, reducing the quark abundance and ultimately producing trace amounts of color-neutral dark baryons. Our main goal is to compute the surviving baryon abundance after this compression.

{\bf Relevant Properties of the Pockets.---} The initial typical pocket size, its wall velocity (i.e. the pocket contraction rate), and the initial density of quarks therein affect the surviving baryon abundance. 
For simplicity, we assume that all pockets have roughly the same properties. 
In the companion work \cite{bigpaper}, we derive expressions for these quantities and study their effects on the relic abundance calculation in more detail. 
Here, we merely summarize our final estimates for these quantities.

\textit{Pocket wall velocity:} Bubble nucleation and expansion occurs when the universe supercools slightly below $T_c$ -- at $T_c$ phase conversion is impossible.
As a bubble's wall expands, deconfined phase converts to confined phase, liberating latent heat near the wall. 
For the confinement phase transition, the latent heat is large enough~\cite{Lucini:2005vg} to heat the wall back up to $T_c$, thus impeding the motion of the wall. 
The wall velocity is therefore limited by the rate at which heat flows away from the wall \cite{Witten:1984rs}. 
A similar argument holds for the pocket contraction rate. 
Immediately after percolation, sometime between the middle and the right top plots in Fig.~\ref{fig:phases}, we model the heat flow with a heat diffusion equation with characteristic length scale $\Lambda^{-1}$ and assume that the pocket has attained a steady state at which this heat diffusion exactly balances the rate at which latent heat is injected. As the pocket contracts, the compressed quarks build up a pressure that counters the pocket contraction and further slows its wall velocity. While we can obtain an expression for the pocket wall velocity $v_w$ when the quark pressure is negligible, a better understanding of the underlying strong dynamics is required when the quark pressure effects are included. However, as we will explain shortly, this does not pose an obstacle in finding the viable parameter space of our model.

\textit{Initial pocket size: }At the time of percolation, the bubbles of the confined phase come into contact with one another and begin to merge. 
As bubbles merge and grow in size, the merger rate slows down, since it takes more time to move more matter. 
At some critical size, $R_1$, the timescale for further merging exceeds the phase transition timescale. 
This critical size determines the typical size of the pockets after percolation, which, following the analysis in Ref.~\cite{Witten:1984rs}, we find to be
\begin{equation}
    R_1 \approx \left( \frac{M_{\mathrm{Pl}}}{10^4\Lambda} \right)^{2/3} \frac{1}{\Lambda},
    \label{eq:initial_R}
\end{equation}
where $M_{\mathrm{pl}}$ is the reduced Planck mass.

\textit{Initial density of the quarks: }
The density of the quarks at the onset of the phase transition can be determined from their freezeout abundance, and is given by 
$N_q^{\mathrm{initial}}  =  (4\pi/3) \, R_1^3 \, n_q$, with $n_q$ the quark number density at the onset of the phase transition. We determine $n_q$ by numerically solving the Boltzmann equations governing quark freezeout.

These are all the quantities we need from the complicated dynamics of the bubbles and the pockets. In Ref.~\cite{bigpaper} we provide a more detailed discussion of the approximations used above and show how they enter the relic abundance calculation.

{\bf Local Recoupling and Baryon Abundance.---}
During the pocket contraction epoch, bound states are formed, in a chain of reactions culminating in color-neutral baryons that can escape the pocket. Simultaneously, much of the initial quark abundance annihilates away. There will be a statistical over-abundance of quarks or anti-quarks with expectation value $\sqrt{N_q^{\mathrm{initial}}}$ in each pocket, resulting in a local matter-antimatter asymmetry in the dark sector. This places an upper bound on the fraction of DM that can annihilate; in the limit of fast annihilation, essentially all of the under-abundant species (quarks or anti-quarks) will annihilate away, leaving a residue of the over-abundant species. The final dark baryon yield from the pocket will then be controlled solely by the local asymmetry, which in turn is determined by the size of the pockets when they are formed. We refer to this simple limiting scenario as {\it accidentally asymmetric} dark matter.

More generally, we define a survival factor $\mathcal{S}$ as the ratio of the quarks + antiquarks that survive within baryons by the end of the phase transition, to the initial number of free quarks + antiquarks. If the symmetric abundance in a contracting pocket is reduced by more than $1/\sqrt{N_q^{\mathrm{initial}}}$, the stochastic asymmetric abundance will constitute the dominant abundance of dark baryons at the end of the phase transition, and the accidentally asymmetric limit will be a good approximation.

In the remainder of this section, we build the tools to determine whether this asymmetric limit is realized, by considering the evolution of the symmetric abundance during the pocket contraction epoch. 
The degrees of freedom in the pockets are the free quarks, diquarks, baryons, and gluons. We neglect bound states like $\bar{q}q$ mesons and more exotic hadrons as they promptly decay into simpler bound states or, through an unspecified portal, to the SM. 

In what follows we denote each particle by its quark number. We can use the conservation of the quark number to write down all the possible 2-to-2 processes that enter the following set of Boltzmann equations, which determine the abundances of each particle state during the compression:
\begin{equation}
 \label{eq:fullboltz}
     L[i] = -\!\!\!\!\sum_{ a+b=c+d} \!\!\!\! s^i_{a,b,c,d}\langle \sigma v \rangle_{ab \rightarrow cd} \left(	n_a n_b - n_c n_d \frac{n_a^{eq} n_b^{eq}}{n_c^{eq} n_d^{eq}}	\right)
\end{equation}
for $i=1,2,3$ and where $s^i_{a,b,c,d}$ is the net number of $i$ particles destroyed in the $ab \rightarrow cd$ process, 
and $L[i]$ denotes the Liouville  operator for relic $i$
\begin{eqnarray}
L[i] &=& \dot{n}_i - 3 \frac{v_w}{R} n_i, ~~~ i=1,2\label{eq:L12}\\
    L[3] &=& \dot{n}_3 + 3 \frac{ v_q }{R} n_3.
\label{eq:L3}
\end{eqnarray}
Here $\langle \sigma v \rangle$ is the cross section of the relevant process, $n_\alpha$ is the number density of species $\alpha$ {\it inside the pocket} while $n^{eq}_\alpha$ is its value in equilibrium, $R$ is the pocket radius, and $v_q \simeq \sqrt{\Lambda/m_q}$ is the typical quark velocity. Finally, the number of baryons that escape the pocket is determined by
\begin{equation}
    dN_3^{\mathrm{esc}} = 4 \pi R^2 n_3 (R)(v_q+v_w) \, dt 
\label{eq:baryonesc}
\end{equation}

The Liouville operator in Eq.~\eqref{eq:L12} decribes a particle confined to a contracting pocket of radius $R$ and wall velocity $v_w$. On the other hand, the baryon Liouville operator in Eq.~\eqref{eq:L3} does not include any contraction terms since the baryons do not feel the presence of the pocket wall. They form inside the pocket, but can subsequently move freely into the confined phase.

The survival factor $\mathcal{S}$, assuming quark-antiquark symmetry, is then given by,
\begin{equation}
    \mathcal{S} = \frac{3\int dN_3^{\mathrm{esc}} }{N_q^{\mathrm{initial}}},
\label{eq:final-Sfactor-symm}
\end{equation}
where the integral is taken over the entire pocket-contraction time. 
In Ref.~\cite{bigpaper} we use the solutions of the Boltzmann equations in Eq.~\eqref{eq:fullboltz} to calculate this quantity for different parts of the parameter space. 
We also derive an asymptotic, analytic expression for this quantity 
\begin{align}
\label{eq:Sanalitic}
\mathcal{S} = 9 \frac{v_q}{v_w} \frac{(N_q^{\mathrm{initial}})^2}{\tilde{f}_1^2 V_{\rm rec}^2}\approx 9 \frac{v_q}{v_w} \frac{4\pi  v_w^3  }{3 \tilde{f}_1^2 N_q^{\mathrm{initial}} \langle \sigma v  \rangle^3_{1(-1)\rightarrow 00}}\, ,
\end{align} 
where $\tilde{f}_1 \equiv \frac{(n_1^{eq})^2}{n_2^{eq}} \sim (m_q T_c)^{3/2}\exp{(- \Delta E/T_c)}$ with $\Delta E$ denoting the heat released during the diquark production process. 
This expression provides us with an intuitive interpretation of the survival factor. Increasing the quark velocity $v_q$ enhances their escape rate, thus increasing $\Sfactor$. Increasing $\langle \sigma v \rangle_{1(-1)\rightarrow 00}$ also decreases the survival factor; this is expected, since by increasing this cross section quarks annihilate more against each other, instead of binding in bound states.
We also find that the survival factor decreases as the initial number of quarks in the pocket increases. The initial number of quarks in the pocket, in turn, is a function of the initial quark density in the pocket and the pocket's initial radius.

Combining this result for the symmetric component with the accidental asymmetric contribution discussed earlier, the baryon survival factor at the end of the phase transition is,
\begin{align}
\label{eq:S-final}
\mathcal{S} = \max{ \left(\frac{3\int dN_3^{\mathrm{esc}}}{N_q^{\mathrm{initial}}},  \frac{\sqrt{N_q^{\mathrm{initial}}}}{N_q^{\mathrm{initial}}}\right)}\,.
\end{align}

This equation is our main result, and is in stark contrast to Ref.~\cite{1707.05380}, where only an $\mathcal{O}(1)$ combinatoric suppression was considered. We find that, while valid in some regimes of the parameter space, Ref.~\cite{1707.05380} grossly overestimates the real dark quark abundance in a large fraction of the phenomenologically viable parameter space.

\begin{figure}[t!]
\resizebox{0.9\columnwidth}{!}{
\includegraphics[scale=1]{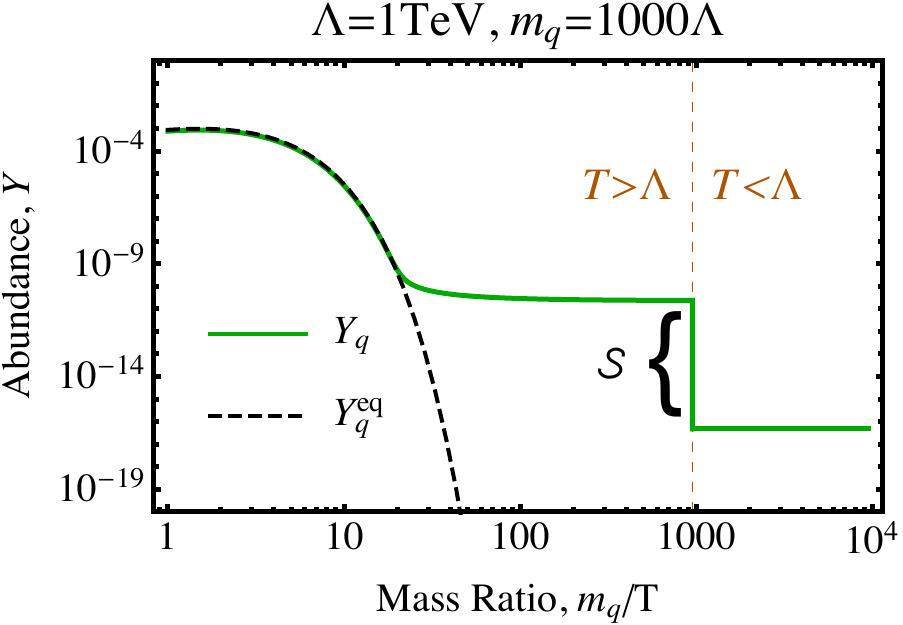}
}
\caption{
The evolution of the DM abundance. In deriving these plots, the quark pressure effect on the wall velocity is neglected. The figure illustrates that only a fraction of all the initial free quarks in the pocket will end up inside stable baryons and survive the new stage of annihilation at $T=\Lambda$. 
}
\label{fig:abundances-R}
\end{figure}

In Fig.~\ref{fig:abundances-R}, we show an example of the evolution of the dark quark abundance, starting from before the phase transition temperature. To obtain a calculable result and build intuition, we set the quark pressure to zero for this example. We find a substantial suppression in the number of quarks due to their compression during the pocket contraction. 

As mentioned before, including the quark pressure slows down the wall velocity, thus reducing the survival factor according to Eq.~\eqref{eq:Sanalitic}. Consequently, a non-zero quark pressure reduces the symmetric survival factor even further, and drives the system toward the accidentally asymmetric regime, where the present-day DM density is set by the local asymmetry of dark quarks in isolated pockets during the dark confinement phase transition.

{\bf Dark Matter Parameter Space.---}
We solve the Boltzmann equations in Eq.~\eqref{eq:fullboltz} for different masses and confinement scales to determine the survival factor. The most severe possible suppression corresponds to saturating the asymmetry bound in Eq.~\eqref{eq:S-final}. 
In fact, we find via a numerical scan~\cite{bigpaper} that even with zero quark pressure $S$ generically comes quite close to saturating this asymmetry bound, so that even a modest quark pressure effect is capable of achieving this saturation over all parameter space. Furthermore, we test the self-consistency of the zero-quark-pressure approximation and find that quark pressure is expected to be non-negligible for all relevant parameter space \cite{bigpaper}. Consequently, we expect the accidentally-asymmetric regime to be a good approximation for the entire parameter space once the quark pressure and any other strong-dynamics effects are carefully taken into account.

\begin{figure}
\resizebox{0.9\columnwidth}{!}{
\includegraphics[scale=1]{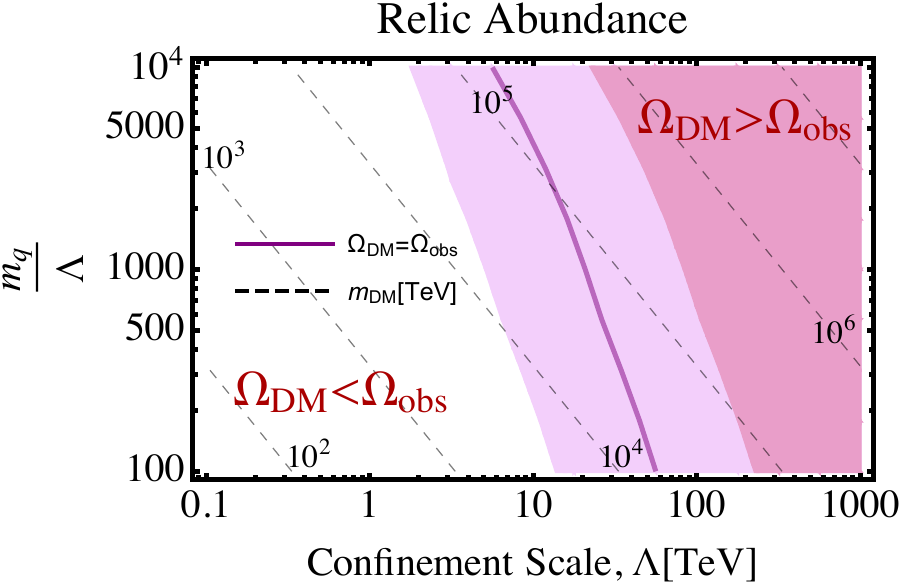}
}
\caption{The dark matter parameter space in this setup. The observed DM abundance is obtained for any point on the purple line. The light purple band shows the effect of increasing/decreasing the initial pocket radius by an order of magnitude. Any point above the relic abundance line produces too much dark matter and is ruled out. The DM relic abundance in the entire region is determined by only the accidental asymmetric abundance in each pocket. The contours of constant mass are shown by dashed black lines.}
\label{fig:phasespace}  
\end{figure}

Under this assumption of accidental asymmetry, Fig.~\ref{fig:phasespace} shows the available parameter space assuming ${\Omega_{\rm DM} = 0.26}$ \cite{1807.06209}. 
It should be noted that the dark quark mass is significantly larger than the mass expected in a purely combinatorial recombination of dark quarks into baryons \cite{1707.05380}, with dark baryons of order $1-100~ \,\rm PeV$ yielding the correct abundance. Such high masses are unexpected for thermally produced DM, as the unitarity bound limits the mass to be $m_{\rm DM} \lesssim  200 \, \rm TeV$ \cite{1407.7874,1904.11503}. Since the asymmetric component of dark quarks is all that survives the pocket contraction, the dark relic abundance is predominantly determined by the initial size of the pockets and the initial quark number trapped therein. 

The initial pocket radius is the main source of uncertainty in our results. To study its effect on the relic abundance, we parametrize the theoretical uncertainty by multiplying the central value of Eq.~\eqref{eq:initial_R} by a multiplicative factor of $(0.1,~10)$. The relic abundance line in Fig.~\ref{fig:phasespace} moves within the light purple band as we vary this factor in this range. We discuss the uncertainty in the relic abundance calculation stemming from various quantities in more details in Ref.~\cite{bigpaper}.

In our calculation, we neglected the initial abundance of the bound states before the confinement. We check that this is justified in the parameter space included in Fig.~\ref{fig:phasespace}; however, we find that this assumption fails as we go to lower values of $\Lambda$ or larger $m_q/\Lambda$. A proper study of these parts of the parameter space requires a more careful study of this initial condition and is left for future works. For lower values of $m_q/\Lambda$, eventually the phase transition ceases to be first-order \cite{1106.0974}, eliminating the second annihilation stage.

{\bf Conclusions.---} A confining dark sector with dark quarks goes through a first-order phase transition when the quarks are heavy enough. In such a dark sector the quarks are trapped inside pockets of the deconfined phase during this phase transition and compressed. Due to this compression, the quarks recouple and go through a new stage of annihilation. Since the process of trapping the quarks inside the pockets is inherently random, each pocket has an accidental asymmetry between the number of quarks and anti-quarks. 
This asymmetry puts a lower bound on the survival factor $\mathcal{S}$ of quarks during the phase transition. We estimated that this lower bound will be saturated in all parts of the studied parameter space.
~\\
Due to the dramatic suppression of the relic abundance during the pocket contraction, the observed DM abundance is explained with very heavy DM masses of order  $ 10^3-10^5$~TeV. Given its low number density, such heavy DM might have escaped detection, despite possibly having a significant interaction strength with SM particles. An experimental detection of such a heavy relic may require a completely novel approach. Deeper theoretical understanding of concrete models and a dedicated search program are needed in order to explore this scenario.

\section*{Acknowledgments}
We are grateful for helpful discussions with Andrei Alexandru, Tom Cohen, Daniel Hackett, Julian Mu$\tilde{\mathrm{n}}$oz, Jessie Shelton, and Xiaojun Yao. We especially thank Benjamin Svetitsky for his collaboration in early stages of this project. We also thank Yonit Hochberg, Graham Kribs, Rebecca Leane, Michele Redi, and Kai Schmidt-Hoberg for constructive comments on the draft. We especially thank Filippo Sala for raising the question of quark pressure effects. 
The work of PA, GWR, and TRS was supported by the U.S. Department of Energy, Office of Science, Office of High Energy Physics, under grant Contract Number DE-SC0012567. 
The work of PA is also supported by the MIT Department of Physics. 
GWR was also supported by an NSF GRFP and the U.S. Department of Energy, Office of Science, Office of Nuclear Physics under grant Contract Number DE-SC0011090. 
The work of EK is supported by the Israel Science Foundation (grant No.1111/17), by the Binational Science Foundation
(grant No. 2016153), and by the I-CORE Program of the Planning Budgeting Committee (grant No. 1937/12). 
The work of EDK was supported by the Zuckerman STEM Leadership Program, and by ISF and I-CORE grants of EK. 
JS is primarily supported by a Feodor Lynen Fellowship from the Alexander von Humboldt foundation.

\footnotesize
\bibliographystyle{abbrv}

\begin{thebibliography}{nnn}\bibitem{hep-ph/0608055}
\article[hep-ph/0608055]{S.B. Gudnason, C. Kouvaris, F. Sannino}{Phys. Rev.}{D74}{095008}{2006}
{Dark Matter from new Technicolor Theories}.


\bibitem{0903.3945}
\article[0903.3945]{D.S.M. Alves, S.R. Behbahani, P. Schuster, J.G. Wacker}{Phys. Lett.}{B692}{323}{2009}
{Composite Inelastic Dark Matter}.


\bibitem{0906.0577}
\article[0906.0577]{C. Kilic, T. Okui, R. Sundrum}{JHEP}{1002}{018}{2009}
{Vectorlike Confinement at the LHC}.


\bibitem{0907.1007}
\article[0907.1007]{T. Hambye, M.H.G. Tytgat}{Phys. Lett.}{B683}{39}{2009}
{Confined hidden vector dark matter}.


\bibitem{0909.2034}
\article[0909.2034]{G.D. Kribs, T.S. Roy, J. Terning, K.M. Zurek}{Phys. Rev.}{D81}{095001}{2009}
{Quirky Composite Dark Matter}.


\bibitem{1003.4729}
\article[1003.4729]{D. Spier Moreira Alves, S.R. Behbahani, P. Schuster, J.G. Wacker}{JHEP}{1006}{113}{2010}
{The Cosmology of Composite Inelastic Dark Matter}.


\bibitem{1005.0008}
\article[1005.0008]{Y. Bai, R.J. Hill}{Phys. Rev.}{D82}{111701}{2010}
{Weakly Interacting Stable Pions}.


\bibitem{1102.0282}
\article[1102.0282]{J.L. Feng, Y. Shadmi}{Phys. Rev.}{D83}{095011}{2011}
{WIMPless Dark Matter from Non-Abelian Hidden Sectors with Anomaly-Mediated Supersymmetry Breaking}.


\bibitem{1106.3101}
\article[1106.3101]{R. Fok, G.D. Kribs}{Phys. Rev.}{D84}{035001}{2011}
{Chiral Quirkonium Decays}.


\bibitem{1109.3513}
\article[1109.3513]{R. Lewis, C. Pica, F. Sannino}{Phys. Rev.}{D85}{014504}{2011}
{Light Asymmetric Dark Matter on the Lattice: SU(2) Technicolor with Two Fundamental Flavors}.


\bibitem{1204.2808}
\article[1204.2808]{M. Frigerio, A. Pomarol, F. Riva, A. Urbano}{JHEP}{1207}{015}{2012}
{Composite Scalar Dark Matter}.


\bibitem{1209.6054}
\article[1209.6054]{M.R. Buckley, E.T. Neil}{Phys. Rev.}{D87}{043510}{2013}
{Thermal dark matter from a confining sector}.


\bibitem{1301.1693}
\article[1301.1693]{{\sc LSD } Collaboration}{Phys. Rev.}{D88}{014502}{2013}
{Lattice Calculation of Composite Dark Matter Form Factors}.


\bibitem{1307.2647}
\article[1307.2647]{S. Bhattacharya, B. Meli{\' c}, J. Wudka}{JHEP}{1402}{115}{2014}
{Pionic Dark Matter}.


\bibitem{1312.3325}
\article[1312.3325]{J.M. Cline, Z. Liu, G. Moore, W. Xue}{Phys. Rev.}{D90}{015023}{2014}
{Composite strongly interacting dark matter}.


\bibitem{1402.3629}
\article[1402.3629]{K.K. Boddy, J.L. Feng, M. Kaplinghat, T.M.P. Tait}{Phys. Rev.}{D89}{115017}{2014}
{Self-Interacting Dark Matter from a Non-Abelian Hidden Sector}.


\bibitem{1408.6532}
\article[1408.6532]{K.K. Boddy, J.L. Feng, M. Kaplinghat, Y. Shadmi, T.M.P. Tait}{Phys. Rev.}{D90}{095016}{2014}
{Strongly interacting dark matter: Self-interactions and keV lines}.


\bibitem{1411.3727}
\article[1411.3727]{Y. Hochberg, E. Kuflik, H. Murayama, T. Volansky, J.G. Wacker}{Phys. Rev. Lett.}{115}{021301}{2015}
{Model for Thermal Relic Dark Matter of Strongly Interacting Massive Particles}.


\bibitem{1503.04203}
\article[1503.04203]{{\sc LSD } Collaboration}{Phys. Rev.}{D92}{075030}{2015}
{Stealth Dark Matter: Dark scalar baryons through the Higgs portal}.


\bibitem{1503.08749}
\article[1503.08749]{O. Antipin, M. Redi, A. Strumia, E. Vigiani}{JHEP}{1507}{039}{2015}
{Accidental Composite Dark Matter}.


\bibitem{1602.00714}
\article[1602.00714]{A. Soni, Y. Zhang}{Phys. Rev.}{D93}{115025}{2016}
{Hidden SU(N) Glueball Dark Matter}.


\bibitem{1604.04627}
\article[1604.04627]{G.D. Kribs, E.T. Neil}{Int. J. Mod. Phys.}{A31}{1643004}{2016}
{Review of strongly-coupled composite dark matter models and lattice simulations}.


\bibitem{1606.00159}
\article[1606.00159]{K. Harigaya, M. Ibe, K. Kaneta, W. Nakano, M. Suzuki}{JHEP}{1608}{151}{2016}
{Thermal Relic Dark Matter Beyond the Unitarity Limit}.


\bibitem{1707.05380}
\article[1707.05380]{A. Mitridate, M. Redi, J. Smirnov, A. Strumia}{JHEP}{1710}{210}{2017}
{Dark Matter as a weakly coupled Dark Baryon}.


\bibitem{1801.01135}
\article[1801.01135]{V. De Luca, A. Mitridate, M. Redi, J. Smirnov, A. Strumia}{Phys. Rev.}{D97}{115024}{2018}
{Colored Dark Matter}.


\bibitem{1802.07720}
\article[1802.07720]{M. Geller, S. Iwamoto, G. Lee, Y. Shadmi, O. Telem}{JHEP}{1806}{135}{2018}
{Dark quarkonium formation in the early universe}.


\bibitem{1811.06975}
\article[1811.06975]{R. Contino, A. Mitridate, A. Podo, M. Redi}{JHEP}{1902}{187}{2019}
{Gluequark Dark Matter}.


\bibitem{1811.08418}
\article[1811.08418]{C. Gross, A. Mitridate, M. Redi, J. Smirnov, A. Strumia}{Phys. Rev.}{D99}{016024}{2019}
{Cosmological Abundance of Colored Relics}.


\bibitem{1904.12013}
\article[1904.12013]{V. Beylin, M.Y. Khlopov, V. Kuksa, N. Volchanskiy}{Symmetry}{11}{587}{2019}
{Hadronic and Hadron-Like Physics of Dark Matter}.


\bibitem{1905.08810}
\article[1905.08810]{N.A. Dondi, F. Sannino, J. Smirnov}{Phys. Rev.}{D101}{103010}{2020}
{Thermal history of composite dark matter}.
 


\bibitem{1911.04502}
\article[1911.04502]{D. Buttazzo, L. Di Luzio, P. Ghorbani, C. Gross, G. Landini, A. Strumia, D. Teresi, J-W. Wang}{JHEP}{2001}{130}{2020}
{Scalar gauge dynamics and Dark Matter}.


\bibitem{2004.03299}
\article[2004.03299]{G. Landini, J-W. Wang}{JHEP}{2006}{167}{2020}
{Dark Matter in scalar Sp($ \mathcal{N} $) gauge dynamics}.


\bibitem{2006.16429}
\article[2006.16429]{{\sc LSD } Collaboration}{Phys. Rev.}{D103}{014505}{2021}
{Stealth dark matter confinement transition and gravitational waves}.


\bibitem{2008.10607}
\heparticle[2008.10607]{R. Contino, A. Podo, F. Revello}{Composite Dark Matter from Strongly-Interacting Chiral Dynamics}.


\bibitem{2010.13678}
\article[2010.13678]{V. Beylin, M. Khlopov, V. Kuksa, N. Volchanskiy}{Universe}{6}{196}{2020}
{New physics of strong interaction and Dark Universe}.


\bibitem{GriestKamionkowski} 
\article[Griest:1989wd]{Kim Griest, Marc Kamionkowski}{\PRL}{64}{1990}{615}
{Unitarity Limits on the Mass and Radius of DM Particles}


\bibitem{1407.7874}
\article[1407.7874]{B. von Harling, K. Petraki}{JCAP}{1412}{033}{2014}
{Bound-state formation for thermal relic dark matter and unitarity}.


\bibitem{1904.11503}
\article[1904.11503]{J. Smirnov, J.F. Beacom}{Phys. Rev.}{D100}{043029}{2019}
{TeV-Scale Thermal WIMPs: Unitarity and its Consequences}.


\bibitem{bigpaper}
\article[arXiv]{Pouya  Asadi, Eric  David  Kramer, Eric  Kuflik, Gregory  W.  Ridgway, Tracy  R.  Slatyer, and  Juri  Smirnov}{}{}{}{}
{Dark Bubbles, Recoupling, and the Unitarity Bound}.


\bibitem{Svetitsky:1982gs}
\article[Svetitsky:1982gs]{B. Svetitsky, L.G. Yaffe}{Nucl. Phys.}{B210}{423}{1982}
{Critical Behavior at Finite Temperature Confinement Transitions}.


\bibitem{Kaczmarek:1999mm}
\article[Kaczmarek:1999mm]{O. Kaczmarek, F. Karsch, E. Laermann, M. Lutgemeier}{Phys. Rev.}{D62}{034021}{1999}
{Heavy quark potentials in quenched QCD at high temperature}.


\bibitem{Alexandrou:1998wv}
\article[Alexandrou:1998wv]{C. Alexandrou, A. Borici, A. Feo, P. de Forcrand, A. Galli, F. Jegerlehner, T. Takaishi}{Phys. Rev.}{D60}{034504}{1998}
{The Deconfinement phase transition in one flavor QCD}.


\bibitem{Aoki:2006we}
\article[Aoki:2006we]{Y. Aoki, G. Endrodi, Z. Fodor, S.D. Katz, K.K. Szabo}{Nature}{443}{675}{2006}
{The Order of the quantum chromodynamics transition predicted by the standard model of particle physics}.


\bibitem{Saito:2011fs}
\article[1106.0974]{H. Saito, S. Ejiri, S. Aoki, T. Hatsuda, K. Kanaya, Y. Maezawa, H. Ohno, T. Umeda}{Phys. Rev.}{D84}{054502}{2011}
{Phase structure of finite temperature QCD in the heavy quark region}.


\bibitem{Witten:1984rs}
\article[Witten:1984rs]{E. Witten}{Phys. Rev.}{D30}{272}{1984}
{Cosmic Separation of Phases}.


\bibitem{0805.4642}
\article[0805.4642]{J. Kang, M.A. Luty}{JHEP}{0911}{065}{2008}
{Macroscopic Strings and 'Quirks' at Colliders}.


\bibitem{Lucini:2005vg}
\article[Lucini:2005vg]{B. Lucini, M. Teper, U. Wenger}{JHEP}{0502}{033}{2005}
{Properties of the deconfining phase transition in SU(N) gauge theories}.


\bibitem{1807.06209}
\article[1807.06209]{Planck Collaboration}{Astron. Astrophys.}{641}{A6}{2020}
{Planck 2018 results. VI. Cosmological parameters}.


\bibitem{1106.0974}
\article[1106.0974]{H. Saito, S. Ejiri, S. Aoki, T. Hatsuda, K. Kanaya, Y. Maezawa, H. Ohno, T. Umeda}{Phys. Rev.}{D84}{054502}{2011}
{Phase structure of finite temperature QCD in the heavy quark region}.


\end{thebibliography}

\end{document}